\documentclass{sig-alternate-2013}
\usepackage{enumerate}
\usepackage{graphicx}
\usepackage{subfigure}
\usepackage{epstopdf}
\usepackage{algorithm}
\usepackage{algorithmicx}
\usepackage{algpseudocode}
\usepackage{paralist}
\usepackage[shortcuts]{extdash}
\usepackage{listings}

\newcommand{\CS}{\mathcal{S}}
\newcommand{\seq}{\mathbf{s}}
\newcommand{\tree}{\mathcal{T}}
\newcommand{\As}{\mathcal{A}(\seq)}
\newcommand{\Ast}{\mathcal{A}(\seq_t)}
\newcommand{\expert}{\mu}
\newcommand{\nodes}{\mathcal{V}}
\newcommand{\edges}{\mathcal{E}}
\newcommand{\news}{\mathcal{N}}
\newcommand \axis {a}

\renewcommand \Pr {\mathop{\mbox{\ensuremath{\mathbb{P}}}}\nolimits}

\newdef{definition}{Definition}

\algnotext{EndFor}
\algnotext{EndIf}

\newfont{\mycrnotice}{ptmr8t at 7pt}
\newfont{\myconfname}{ptmri8t at 7pt}
\let\crnotice\mycrnotice%
\let\confname\myconfname%

\permission{\textbf{A version of this paper appeared in the Proceedings of the 7th ACM conference on Recommender Systems (RecSys 2013), pages 105-112, Hong Kong, China.}}

\def\sharedaffiliation{%
\end{tabular}
\begin{tabular}{c}}

\begin{document}

\newpage
\thispagestyle{empty}

\onecolumn

\vspace*{\fill}
\begingroup
\centering
\textbf{A version of this paper appeared in the Proceedings of the 7th ACM conference on Recommender Systems (RecSys 2013), pages 105-112, Hong Kong, China.}

\textbf{Please cite as} \emph{F. Garcin, C. Dimitrakakis, and Boi Faltings. ``Personalized news recommendation with context trees.'' Proceedings of the 7th ACM conference on Recommender systems. ACM, 2013.}

\begin{lstlisting}
@inproceedings{garcin2013CT,
  title={Personalized news recommendation with context trees},
  author={Garcin, Florent and Dimitrakakis, Christos and Faltings, Boi},
  booktitle={Proceedings of the 7th ACM conference on Recommender systems},
  pages={105--112},
  year={2013},
  organization={ACM}
}
\end{lstlisting}

\endgroup
\vspace*{\fill}

\newpage

\title{Personalized News Recommendation with Context Trees}

\numberofauthors{3}

%\author{
%\alignauthor
%Florent Garcin\\
%\affaddr{Artificial Intelligence Lab}\\
%\affaddr{Ecole Polytechnique F\'ed\'erale de Lausanne}\\
%\affaddr{Switzerland}\\
%\email{\normalsize{florent.garcin@epfl.ch}}
%%
%\alignauthor
%Christos Dimitrakakis\\
%\affaddr{Artificial Intelligence Lab}\\
%\affaddr{Ecole Polytechnique F\'ed\'erale de Lausanne}\\
%\affaddr{Switzerland}\\
%\email{\normalsize{christos.dimitrakakis@epfl.ch}}
%%
%\alignauthor
%Boi Faltings\\
%\affaddr{Artificial Intelligence Lab}\\
%\affaddr{Ecole Polytechnique F\'ed\'erale de Lausanne}\\
%\affaddr{Switzerland}\\
%\email{\normalsize{boi.faltings@epfl.ch}}
%}

\author{
\alignauthor Florent Garcin
\alignauthor Christos Dimitrakakis
\alignauthor Boi Faltings
\sharedaffiliation
\affaddr{Artificial Intelligence Lab}\\
\affaddr{Ecole Polytechnique F\'ed\'erale de Lausanne}\\
\affaddr{Switzerland}\\
\email{\normalsize{firstname.lastname@epfl.ch}}
}

\maketitle

\begin{abstract}
The proliferation of online news creates a need for filtering interesting articles. Compared to other products, however, recommending news has specific challenges: news preferences are subject to trends, users do not want to see multiple articles with similar content, and frequently we have insufficient information to profile the reader.

In this paper, we introduce a class of news recommendation systems based on context trees. They can provide high-quality news recommendations to anonymous visitors based on present browsing behaviour. Using an unbiased testing methodology, we show that they make accurate and novel recommendations, and that they are sufficiently flexible for the challenges of news recommendation.
\end{abstract}



\begin{figure}[t!]
\centering
\includegraphics[scale=0.2]{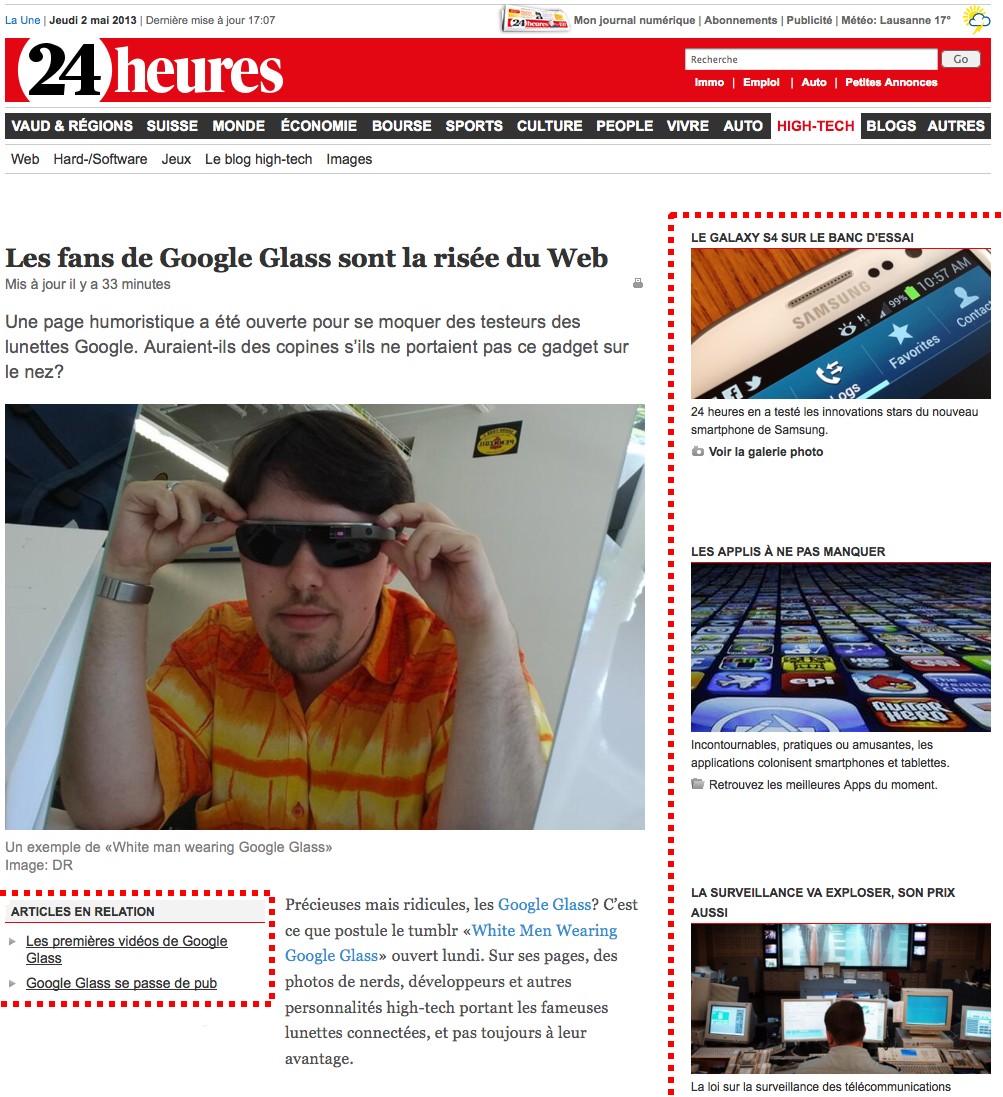}
\caption{A story with dynamic recommendations on the right side, and manually-generated recommendations on the bottom left (red-dashed areas).}
\label{fig:article}
\end{figure}

\section{Introduction}

The first recommender systems were originally designed for news forums. Since then, they have been used with considerable success for products such as books and movies, but have found surprisingly little application in recommending news articles, due to the unique challenges of the area.

When users are identifiable as regular visitors to a news website, techniques from product recommendation can be adapted \cite{Das2007, IJntema2010}. However, most websites operated by individual newspapers do not have a strong base of electronic subscribers. Visitors to these websites are casual users, often accessing them through a search engine, and little is known about them except what can be gathered through an ephemeral browsing history.

The main page of a news website is already a set of recommended articles, which simultaneously addresses the needs of many users (Fig.~\ref{fig:article}). More specific recommendations are sometimes available to readers of individual articles. There are two shortcomings to this strategy: first, recommendations are usually edited manually, and second, they only consider the last article read. Our goal is to construct recommendations automatically and use the complete browsing history as a basis for giving personalized recommendations. 

In principle, common recommender techniques such as collaborative filtering could be applied to such a task, and have been adapted to temporal sequences~\cite{Zimdars2001, Shani2005, Rendle2010}. However, they face several challenges specific to news. First, news are rapidly evolving: new stories and topics appear and disappear quickly, and old news are no longer interesting. Second, recommendations should provide added value, and not just consist of the most popular stories that the reader would have already seen on the front page.

To address these issues, we propose a class of \emph{online} recommendation algorithms based on  \emph{Context Trees (CT)}, which provide recommendations and are updated fully incrementally. A CT defines a \emph{partition tree} organised in a hierarchy of increasingly precise partitions of a space of contexts. We consider as context the sequence of articles, the sequence of topics, or the distribution of topics. Each node in the tree is called \emph{context} and corresponds to a set of sequences or topic distributions within a partition. The main idea is to give context-dependent recommendations, with contexts becoming progressively finer-grained deeper in the tree.

%The choice of space determines which aspect of the user's history we should be basing our recommendations on. In this work, we consider: \emph{a)}  news sequences, \emph{b)} topic sequences, and \emph{c)} topic distributions. The first two (a,b) model the temporal characteristics of users' behaviour. The last model (c) makes recommendations conditional on the distribution of topics preferred by a user. Each of these constructions results in a different behavioural model.

To make actual recommendations, we associate a set of prediction models, called \emph{experts}, with each context. Their predictions are combined to make recommendations. We tailor our expert models to specifically take into account the idiosyncrasies of news. In particular, our expert models take into account the \emph{popularity} and \emph{freshness} of news items.

Using an unbiased testing methodology, emulating the process involved in implementing a system on a real website, we show that CT recommendations have state-of-the-art performance both with respect to prediction accuracy and to recommendation novelty, which is crucial for news articles since users want to read stories they do not know.

\section{Related Work}

In general, there are two classes of recommender systems: collaborative filtering~\cite{Su2009}, which use similar users' preferences to make recommendations, and content-based systems~\cite{Lops2011}, which use content similarity of news items.

The Grouplens project is the earliest example of collaborative filtering for news recommendation, applied to newsgroups \cite{Resnick1994}. News aggregation systems such as Google News \cite{Das2007} also implement such algorithms. Google News uses the probabilistic latent semantic indexing and MinHash for clustering news items, and item covisitation for recommendation. Their system builds a graph where the nodes are the stories and the edges represent the number of covisitations. Each of the approaches generates a score for a given news, aggregated into a single score using a linear combination.

Content-based recommendation is more common for news personalization \cite{Billsus1999, Ahn2007, IJntema2010}. NewsWeeder \cite{Lang1995} is probably the first content-based approach for recommendations, but applied to newsgroups. NewsDude \cite{Billsus1999} and more recently YourNews \cite{Ahn2007} implemented a content-based system.

It is possible to combine the two types in a hybrid system \cite{Burke2002, Liu2010, Li2010}. For example, Liu et al. \cite{Liu2010} extend the Google News study by looking at the user click behaviour in order to create accurate user profiles. They propose a Bayesian model to recommend news based on the user's interests and the news trend of a group of users. They combine this approach with the one by Das et al. \cite{Das2007} to generate personalized recommendations. Li et al. \cite{Li2010} introduce an algorithm based on a contextual bandit which learns to recommend by selecting news stories to serve users based on contextual information about the users and stories. At the same time, the algorithm adapts its selection strategy based on user-click feedback to maximize the total user clicks.

We focus on a class of recommender systems based on context trees. Usually, these trees are used to estimate Variable-order Markov Models (VMM). VMMs have been originally applied to lossless data compression, in which a long sequence of symbols is represented as a set of contexts and statistics about symbols are combined into a predictive model \cite{Rissanen1983}. VMMs have many other applications \cite{Begleiter2004}.

Closely related, variable-order hidden Markov models \cite{Wang2006}, hidden Markov models \cite{Montgomery2004} and Markov models \cite{Pitkow1999, Sarukkai2000, Deshpande2004} have been extensively studied for the related problem of click prediction. These models suffer from high state complexity. Although techniques \cite{Zaki2010} exist to decrease this complexity, multiple models have to be maintained, making these approaches not scalable and not suitable for online learning.

Few works \cite{Zimdars2001, Shani2005, Rendle2010} apply such Markov models to recommender systems. Zimdars et al. \cite{Zimdars2001} describe a sequential model with a fixed history. Predictions are made by learning a forest of decision trees, one for each item. When the number of items is big, this approach does not scale. Shani et al. \cite{Shani2005} consider a finite mixture of Markov models with fixed weights. They need to maintain a reward function in order to solve a Markov decision process for generating recommendations. As future work, they suggest the use of a context-specific mixture of weights to improve prediction accuracy. In this work, we follow such an approach. Rendle et al. \cite{Rendle2010} combine matrix factorization and a Markov chain model for baskets recommendation. The idea of factoring Markov chains is interesting and could be complementary to our approach. Their limitation is that they consider only first-order Markov chains. A bigger order is not tractable because the states are baskets which contain many items.

%These methods rely on a static model that ignores the particular properties of news. Our approach is better suited for such dynamic domains, as the context tree evolves over time and adapts itself to current trends and reader preferences. Classic recommender system approaches such as collaborative filtering require recomputing the model every time. Our algorithm is fully incremental and adapts the model continuously such as \cite{Rendle2008, Liu2010online}. Our approach requires only one tree (the context tree), and thus scales very well. Our work is not restricted to the history of logged-in users, but considers a one-time session for recommendation, where users do not log in. Surprisingly, we do not know of any existing research that considers context-tree models for recommender systems.

\section{Preliminaries}

Because of the sequential nature of news reading, it is intuitive to model news browsing as a $k$-order Markov process \cite{Shani2005}. The user's state can be summarised by the last $k$ items visited, and predictions can be based only on this information. Unfortunately, it is not clear how to select the order $k$. A \emph{variable-order Markov model} (VMM) alleviates this problem by using a context-dependent order. In fact, VMM is a special type of context-tree model~\cite{Begleiter2004}.

There are two key ideas behind a CT recommender system. First, it creates a hierarchy of contexts, arranged in a tree such that a child node completely contains the context of its parents. In this work, a context can be the set of sequences of news items, sequence of topics, or a set of topic distributions. As new articles are added, more contexts are created. Contexts corresponding to old articles are removed as soon as they disappear from the current article pool.

The second key idea is to assign a local prediction model to each context, called an expert. For instance, a particular expert gives predictions only for users who have read a particular sequence of stories, or users who have read an article that was sufficiently close to a particular topic distribution.

In the following, we first introduce the notion of context tree. Then, we describe various prediction models, how to associate them with the context tree and combine them in order to make recommendations.

\subsection{Sequence Context Tree}\label{sec:CT}

When a user browses a news website, we track the sequence of articles read.
\begin{definition}
A \emph{sequence} $\seq =  \langle n_1, \ldots, n_l  \rangle$ is an ordered list of articles $n_i \in \news$ read by a user, and we denote $\seq_t$ the sequence of articles read until time $t$. We write the set of all sequences by $\CS$.
\end{definition}
Note that a sequence can also be a sequence of topics of articles.

A context tree is built based on these sequences and their corresponding suffixes.

\begin{definition}
A $k$-length sequence $\xi$ of is a \emph{suffix} of a $l$-length sequence $\seq$, if $k \leq l$, and the last elements of $\seq$ are equal to $\xi$, and we write $ \xi \prec \seq$ when $\xi$ is a suffix of $\seq$.
\end{definition}
For instance, one suffix $\xi$ of the sequence $\seq = \langle n_1, n_2, n_3, n_4 \rangle$ is given by $\xi = \langle n_3, n_4 \rangle$.

If two sequences have similar context, the next article a user wants to read should also be similar.
\begin{definition}
A \emph{context} $S = \{\seq \in \CS : \xi \prec \seq\}$, $S \subset \CS$ is the set of all possible sequences $\CS$ ending with the suffix $\xi$.
\end{definition}

We can now give a formal definition of a context tree.
\begin{definition}
A \emph{context tree} $\tree = (\nodes, \edges)$ with nodes $\nodes$ and edges $\edges$ is a partition tree over the contexts $\CS$. It has the following properties:
\begin{inparaenum}[(a)]
\item \emph{The set of contexts at a given depth forms a partition}: If $\nodes_k$ are the nodes at depth $k$ of the tree, then $S_i \cap S_j = \emptyset$  $\forall i, j \in \nodes_k$, while $\bigcup_{i \in \nodes_k} S_i = \CS$
\item \emph{Successive refinement}: If node $i$ is the parent of $j$ then $S_j \subset S_i$.
\end{inparaenum}
\end{definition}

\begin{figure}[t!]
\centering
\includegraphics[scale=0.3]{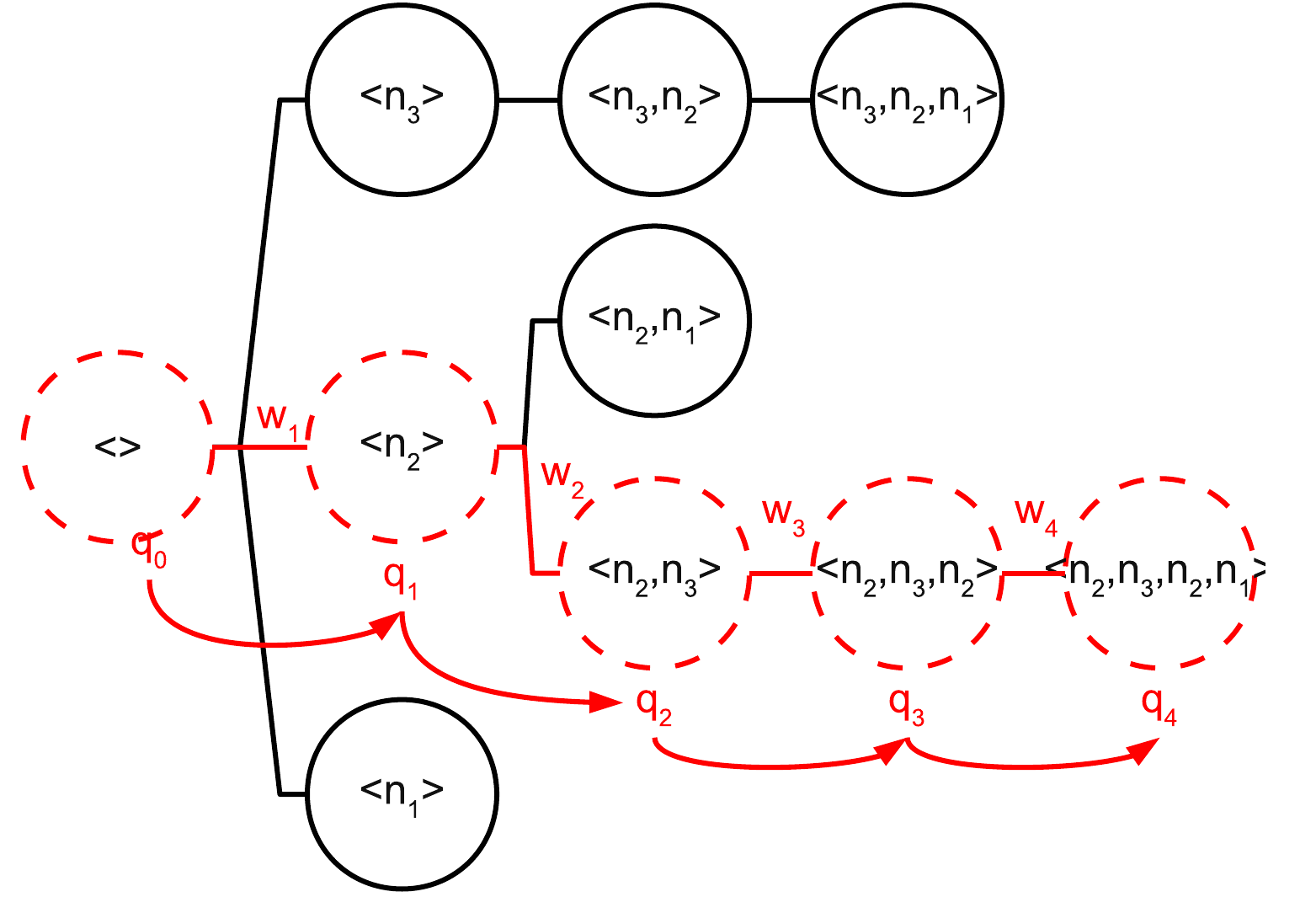}
\caption{VMM context tree for the sequence $\seq = \langle n_1, n_2, n_3, n_2 \rangle$. Nodes in red-dashed are active experts $\mu \in A(\seq)$.}
\label{fig:contexttree}
\end{figure}

Thus, each node $i \in \nodes$ in the context tree corresponds to a context $S_i$. Initially the context tree $\tree$ only contains a root node with context $S_0 = \CS$. Every time a new article $n_t$ is read, the active leaf node is split in a number of subsets, which then become nodes in the tree. This construction results in a variable-order Markov model, illustrated in Fig.~\ref{fig:contexttree}.

The main difference between news articles and products is that articles continuously appear and disappear, and the system thus maintains a current article pool that is always changing. The model for recommendation changes along with the article pool, using a dynamically evolving context tree. As new articles are added, new branches are created corresponding to sequences or topic distributions. At the same time, nodes corresponding to old articles are removed as soon as they disappear from the current pool.

\subsection{Topic Distribution Context Tree}\label{sec:kdCT}

Because of the large number of news items relative to topics, a context tree on topics might make better predictions. In particular, stories that have not been read by anyone can be recommended thanks to topic similarity. In this type of tree, each context represents a subset of the possible \emph{topic distributions} of the last read article. The structure of the tree is slightly different and is modelled via a $k$-d tree.

A $k$-d tree is a binary tree that iteratively partitions a $k$-dimensional space $\CS$ into smaller sets \cite{Bentley1975}. The $i$-th node corresponds to a hyper-rectangle $S_i \subset \CS$ and has two children $j, j'$ such that $S_j \cup S_{j'} = S_i$ and $S_j \cap S_{j'} = \emptyset$. In particular, the two children are always defined via a hyperplane splitting $S_i$ in half, through the center of $S_i$, and which is perpendicular to one principal axis. In practice, we simply associate each node to one of the $k$ axes based on the depth such that we cycle through all possible axes: $\axis = depth~\mathrm{mod}~k$. The set $\CS$ is $[0,1]^{k}$, the set of $k$-dimensional multinomial distributions on the possible topics.

In analogy to the sequence CT, a context is a hyper-rectangle $S_i$ and a suffix is the center $\theta$ of $S_i$ in a topic distribution CT.

For instance, consider a node $i$ with center $\theta \in S_i$ and associated axis $\axis$. Its two children correspond to two sets of topic distributions: Its left child $j$ contains the distributions $\theta' \in S_i$ with $\theta'_\axis < \theta_\axis$, while
its right child $j'$ is the set on the other side of the hyperplane: $S_{j'} = \{\theta' \in S_i : \theta'_\axis \geq \theta_\axis\}$. When the system observes a new topic distribution $\theta$, the distribution is added to the tree, and possibly the tree expands.

\subsection{Experts}

We assign a local prediction model called \emph{expert} to each context (node) in the tree. More formally,
\begin{definition}
An \emph{expert} $\expert_i$ is a function associated with a specific context $S_i$ that computes an estimated probability of the next article $n_{t + 1}$ given that context, i.e. $\Pr_i(n_{t + 1} \mid \seq_t)$.
\end{definition}

The user's browsing history $\seq_t$ is matched to the context tree and identifies a path of nodes (see Fig.~\ref{fig:contexttree}). All experts associated with these nodes are called \emph{active} and are responsible for the recommendation.
\begin{definition}
The set of \emph{active} experts $\Ast = \{ \expert_i : \xi_i \prec \seq_t \}$ is the set of experts $\expert_i$ associated to contexts $S_i = \{\seq : \xi_i \prec \seq_t\}$ such that $\xi_i$ are suffix of $\seq_t$.
\end{definition}

\subsection{Combining Experts into Predictions}

The active experts $\Ast$ are combined by marginalizing to obtain a mixture of probabilities of all active experts:
\begin{equation}\label{eq:mixture}
\Pr(n_{t+1} = x \mid \seq_t) = \sum_{i \in \Ast} u_i(\seq_t) \Pr_i(n_{t+1} = x \mid \seq_t),
\end{equation}
with $u_i(\seq_t) = \Pr(i | \seq_t)$ being the probability of the $i$-th expert relevant for this context. These probabilities are derived as follows. 

With each node $i$ in the context tree we associate a weight $w_i \in [0,1]$ that represents the usefulness of the corresponding expert. Given a path in the context tree, we consider experts in the order of the most specific to the most general context, i.e. along the path from the most specific node to the root. In this process, with probability equal to the weight $w_i$ we stop at a node without considering the more general experts. Thus, we take into account the relative usefulness of the experts. 

Letting $w_j$ be the probability of stopping at $j$ given that we have not stopped yet, we thus obtain the probability $u_i$ that the $i$-th expert is considered as 
$u_i(\seq_t) = w_i \prod_{j : S_j \subset S_i} (1 - w_j)$ if $\seq_t \in S_i$ and $0$ otherwise.

The calculation of the total probability can be made efficiently via the recursion $q_k = w_k \mathbb{P}_k(n_{t+1} = x | \seq_t) + (1 - w_k)q_{k - 1}$, where $q_k$ is the combined prediction of the first $k$ experts. In Figure~\ref{fig:contexttree}, the prediction of the root expert for the next item $x$ is $q_0$, while $q_4$ is the complete prediction by the model for this sequence.

The weights are updated by taking into account the success of a recommendation. When a user reads a new article $x$, we update the weights of the active experts corresponding to the suffixes ending before $x$ according to the probability $q_k(x)$ of predicting $x$ sequentially via Bayes' theorem~\cite{dimitrakakis2010}:
\begin{equation}
w_k' = \frac{w_k \mathbb{P}_k(n_{t+1} = x \mid \seq_t)}{q_k(x)}.
\end{equation}
No other weights are updated\footnote{$w_0 = 1$ since we must always stop at the root node.}. Finally, we also update the local models of the active experts (see Section~\ref{sec:expertmodels}).

\subsection{Expert Models}\label{sec:expertmodels}

Recommending news articles depends on multiple factors: the popularity of the news item, the freshness of the story, the sequence of news items or topics that the user has seen so far. Thus, each expert is decomposed into a set of \emph{local models}, each modelling one of these properties. The first model ignores the temporal dynamics of the process. The second model assumes that users are mainly looking at popular items, and the last model that they are interested in fresh items (i.e. breaking news).

\subsubsection{Standard model}

A na\"{i}ve approach for estimating the multinomial probability distribution over the news items is to use a Dirichlet-multinomial prior for each expert $\mu_i$. The probability of reading a particular news item $x$ depends only on the number of times $\alpha_x$ it has been read when the expert is active.
\begin{equation}\label{eq:std}
\mathbb{P}^{std}_i(n_{t+1} = x | \seq_t) = \frac{\alpha_x + \alpha_0}{\sum_{j \in \news} (\alpha_j + \alpha_0)},
\end{equation}
where $\alpha_0$ is the initial count of the Dirichlet prior.

The dynamic of news items is more complex. A news item provides new content and therefore has been seen by few users. News is subject to trends and frequent variations of preferences. We improve this simple model by augmenting it with models for \emph{popular} or \emph{fresh} news items.

\subsubsection{Popularity model}

A news item $x \in \mathcal{P}$ is in the set of popular items $\mathcal{P}$ when it has been read at least once among the last $|\mathcal{P}|$ read news items. We compute the probability of a news item $x$ given that $x$ is \emph{popular} as:
\begin{equation}
\mathbb{P}^{pop}_i(n_{t+1} = x | \seq_t) = \frac{c_x + \alpha_0}{\sum_{j \in \news} (c_j + \alpha_0)},
\end{equation}
where $c_x$ is the total number of clicks received for news item $x$. Note that $c_x $ is not equal to $\alpha_x$ (Eq. \ref{eq:std}). $\alpha_x$ is the number of clicks for news item $x$ when the expert is \emph{active}, while $c_x$ is the number of clicks received by news item $x$ in \emph{total} whether the expert is active or not.

The number of popular items $|\mathcal{P}|$ is important because it is unique for each news website. When $|\mathcal{P}|$ is small, the expert considers only the most recent read news. It is important to tune this parameter appropriately.

\subsubsection{Freshness model}

A news item $x \in \mathcal{F}$ is in the set of fresh items $\mathcal{F}$ when it has not been read by anyone but is among the next $|\mathcal{F}|$ news items to be published on the website, i.e. a breaking news. We compute the probability of news item $x$ given that $x$ is \emph{fresh} as:
\begin{equation}
\mathbb{P}^{fresh}_i(n_{t+1} = x | \seq_t) = \left\{
\begin{array}{cl}
\frac{1}{|\mathcal{F}| + 1}, & \text{if } x \in \mathcal{F}\\
\frac{1}{(|\mathcal{F}| + 1)(|\news| - |\mathcal{F}|)}, & \text{if } x \notin \mathcal{F}.
\end{array}\right.
\end{equation}

The number of fresh items $|\mathcal{F}|$ influences the prediction made by this expert, and is unique for each news website.

\subsubsection{Mixing the expert models}

We combine the three expert models using this mixture:
\begin{equation}
\mathbb{P}_i(n_{t+1} = x | \seq_t) = \sum_{\tau \in \{std, pop, fresh\}} \mathbb{P}^{\tau}_i(n_{t+1} = x | \seq_t) p^{\tau}_i.
\end{equation}

There are two ways to compute the probabilities $p^{\tau}_i$: either by using a Dirichlet prior that ignores the expert prediction or by a Bayesian update to calculate the posterior probability of each expert according to their accuracy.

For the first approach, the probability of the next news item being \emph{popular} is:
\begin{eqnarray}\label{eq:update_pop}
p^{pop}_i = \mathbb{P}_i(n_{t+1} \in \mathcal{P}) &=& \frac{\alpha_{pop} + \alpha_0}{(\alpha_{pop} + \alpha_0) + (\alpha_{not pop} + \alpha_0)} \nonumber\\
&=& \frac{\alpha_{pop} + \alpha_0}{2\alpha_0 + \sum_j \alpha_j},
\end{eqnarray}
where $\sum_j \alpha_j$ represents the number of times the expert $\mu_i$ has been active, $\alpha_{pop}$ and $\alpha_{not pop}$ the number of read news items which were respectively popular and not popular when the expert $\mu_i$ was active.

Similarly, the probability of the next news item being \emph{fresh} is given by:
\begin{equation}\label{eq:update_fresh}
p^{fresh}_i = \mathbb{P}_i(n_{t+1} \in \mathcal{F}) = \frac{\alpha_{fresh} + \alpha_0}{2\alpha_0 + \sum_j \alpha_j},
\end{equation}
where $\alpha_{fresh}$ is the number of read news items which were fresh when the expert $\mu_i$ was active.

Noting that $\mathcal{P} \cap \mathcal{F} = \emptyset$, the probability of the next news item being neither popular nor fresh is:
\begin{equation}\label{eq:update_std}
p^{std}_i = \mathbb{P}_i(n_{t+1} \notin \mathcal{P} \cup \mathcal{F}) = 1 - \mathbb{P}_i(n_{t+1} \in \mathcal{P}) - \mathbb{P}_i(n_{t+1} \in \mathcal{F}).
\end{equation}

It might happen that by using the Dirichlet priors, predictions are mainly made by only one expert model. To overcome this issue, we compute the probabilities $p^{\tau}_i$, $\tau \in \{std, pop, fresh\}$ via a Bayesian update, which adapts them based on the performance of each expert model:
\begin{equation}\label{eq:bayesian}
p^{\tau}_i \gets \frac{\mathbb{P}^{\tau}_i(n_{t+1} = x | \seq_t) p^{\tau}_i}{\mathbb{P}_i(n_{t+1} = x | \seq_t)}.
\end{equation}

\section{Context-Tree Recommenders}

We describe here the general algorithm to generate recommendations for the class of context-tree recommender systems. This algorithm can be applied to domains other than news in which timeliness and concept drift are of concern. We then focus on the news domain and describe in more details three VMM-based recommender systems and one based on the $k$-d context tree.

\subsection{General Algorithm}

Algorithm~\ref{alg:recsys} presents a sketch of the CT recommender algorithm. For simplicity, we split our system in two procedures: \emph{learn} and \emph{recommend}. Both are executed for each read article $x$ of a user with browsing history $\seq$ in an online algorithm such as \cite{Rendle2008, Liu2010online}, without any further offline computation. The candidate pool $\mathcal{C}$ is always changing and contains the popular $\mathcal{P}$ and fresh $\mathcal{F}$ stories. The system estimates the probability of each candidate and recommends the news items with the highest probability. In order to estimate the probability of a candidate item, the system \emph{1)} selects the active experts $\As$ which correspond to a path in the context tree from the most general to the most specific context, \emph{2)} propagates $q$ from the root down to the leaf, i.e the most specific context. $q$ at the leaf expert is the estimated probability of the recommender system for the candidate item $x$, i.e. $\mathbb{P}(n_{t+1} = x | \seq_t)$ (see Eq.~\ref{eq:mixture}).

\begin{algorithm}
\caption{CT recommender system}\label{alg:recsys}
\begin{algorithmic}[1]
\Procedure{Learn}{$x, \seq$, context set $\Xi$}
\State $q \gets 0$ and $t \gets |\As|$
\item[\Comment{{\small loop from most general expert $\mu_0$ to most specific expert $\mu_t$}}]
\For{$i \gets 0, t$}
	\State $p_i \gets \mathbb{P}_i(n_{t+1} = x | \seq)$ \Comment{{\small $i$th expert prediction}}
	\State $q \gets w_i p_i + (1 - w_i) q$ \Comment{{\small combined prediction}}
	\State $w_i \gets \frac{w_i p_i}{q}$ \Comment{{\small weight update}}
	\State update $p^{std}_i, p^{pop}_i, p^{fresh}_i$ (Eq.~\ref{eq:update_pop}-\ref{eq:update_std} or Eq.~\ref{eq:bayesian})
%	\State update $\alpha_x, \alpha_{pop}, c_x, \alpha_{fresh}$ according to $x$
\EndFor
\If{$x~\circ~\seq \notin \Xi$} \Comment{{\small is the context in the tree?}}
	\State $\Xi = \Xi \bigcup \{x~\circ~\seq\}$ \Comment{{\small add a new leaf.}}
\EndIf
\EndProcedure

\Procedure{Recommend}{$\seq$}
	\ForAll{candidate $n \in \mathcal{C}$}
		\State $q^{(n)} \gets 0$ and $t \gets |\As|$
		\item[\Comment{{\small loop from most general expert $\mu_0$ to most specific expert $\mu_t$}}]
		\For{$i \gets 0, t$}
			\State $q^{(n)} \gets w_i \mathbb{P}_i(n| \seq) + (1 - w_i) q^{(n)}$
		\EndFor
	\EndFor
	\State $\mathcal{R} \gets$ sort all $n \in \mathcal{C}$ by $q^{(n)}$ in descending order
	\State \Return first $k$ elements of $\mathcal{R}$
\EndProcedure
\end{algorithmic}
\end{algorithm}

\subsection{Recommender Systems}

We consider three VMM variants of recommender systems, and one based on the $k$-d context tree.

\subsubsection{VMM Recsys}

The standard VMM recommender builds a context tree on the sequences of news items. That is $\seq_t = \langle n_1, \ldots, n_t \rangle$ is a sequence of news items, and each active expert predicts $n_{t+1}$, the next news item.

\subsubsection{Content-based VMM (CVMM) Recsys}

In order to build a CT on topic sequences, we find a set of topics for each story, and assign the most probable topic to the news item. We then perform predictions on \emph{topics}.

More precisely, we use the Latent Dirichlet Allocation (LDA) \cite{Blei2003} as a probabilistic topic model. After concatenating the title, summary and content of the news item together, we tokenize the words and remove stopwords. We then apply LDA to all the news stories in the dataset, and obtain a topic distribution vector $\theta^{(n)}$ for each news item $n$.

We now define a context tree as follows. Let $z_t$ be the most probable topic of the $t$-th news item. Then the \emph{topic sequence} is $\seq_t = \langle z_1, ..., z_t \rangle$ and $\xi$ is a suffix of topic sequences. The context tree generates a topic probability distribution $\Pr(z_{t+1} = j | \seq_t)$, while the LDA model provides us with a topic distribution $\Pr(z = j | n)$ for each news item $n$. These are then combined into the following score:
\begin{equation}
score(n \mid \seq_t) = \max_{j}\{\mathbb{P}(z_{t+1} = j \mid \seq_t) \cdot \mathbb{P}(z = j \mid n)\}.
\end{equation}
The system recommends the articles with the highest scores.

\subsubsection{Hybrid VMM (HVMM) Recsys}

We combine the standard VMM with the content-based system into a hybrid version. The context tree is built on \emph{topics}, similarly to the CVMM system, but the experts make predictions about \emph{news items}, like the VMM system.

HVMM system builds a tree in the space of \emph{topic sequences}. Each suffix $\xi$ of size $k$ is a sequence of most probable topics $\langle z_1, z_2, ..., z_k \rangle$. However, all predictive probabilities (Eq.~\ref{eq:mixture} and later) are defined on the space of news items.

\subsubsection{$k$-d Context-Tree ($k$-CT) Recsys}

The CVMM and HVMM structures make predictions on the basis of the sequence of most probable topics. Instead, we consider a model that takes advantage of the complete topic distribution of the last news item. We use a $k$-d tree to build a context model in the space of \emph{topic distributions}.

\subsubsection{Baselines}

In addition, we have the following baselines:
\begin{description}\addtolength{\itemsep}{-0.8\baselineskip}
\item[Z-$k$] is a fixed $k$-order Markov chain recommender similar to the ones by Zimdars et al. \cite{Zimdars2001}.
\item[MinHash] is the minhash system used in Google News \cite{Das2007}.
\item[MostPopular] recommends a set of stories with the highest number of clicks among the last read news items.
\end{description}

\section{Evaluation and Comparison}\label{sec:evalcomp}

We investigate whether the class of CT recommender systems has an advantage over standard methods and if so, what is the best combination of partition and expert model. 

We measure performance both with respect to accuracy and novelty of recommendations. Novelty is essential because it exposes the reader to relevant news items that she would not have seen by herself. Obvious but accurate recommendations of most-popular items are of little use.

We evaluate our systems on two datasets described below. We examine on the first dataset the sensitivity of the CT models to hyperparameters. The second dataset is used to perform an unbiased comparison between the different models. We select the optimal hyperparameters on the first dataset, and then measure the performance on the second dataset. This methodology \cite{Bengio2005} mirrors the approach that would be followed by a practitioner who wants to implement a recommender system on a newspaper website.

\subsection{Datasets}

We collected data from the websites of two daily Swiss-French newspapers called \emph{Tribune de Gen\`{e}ve} (TDG) and \emph{24 Heures} (24H)\footnote{www.tdg.ch and www.24heures.ch}. Their websites contain news stories ranging from local news, national and international events, sports to culture and entertainment.

\begin{figure}[t!]
\centering
\includegraphics[scale=0.5]{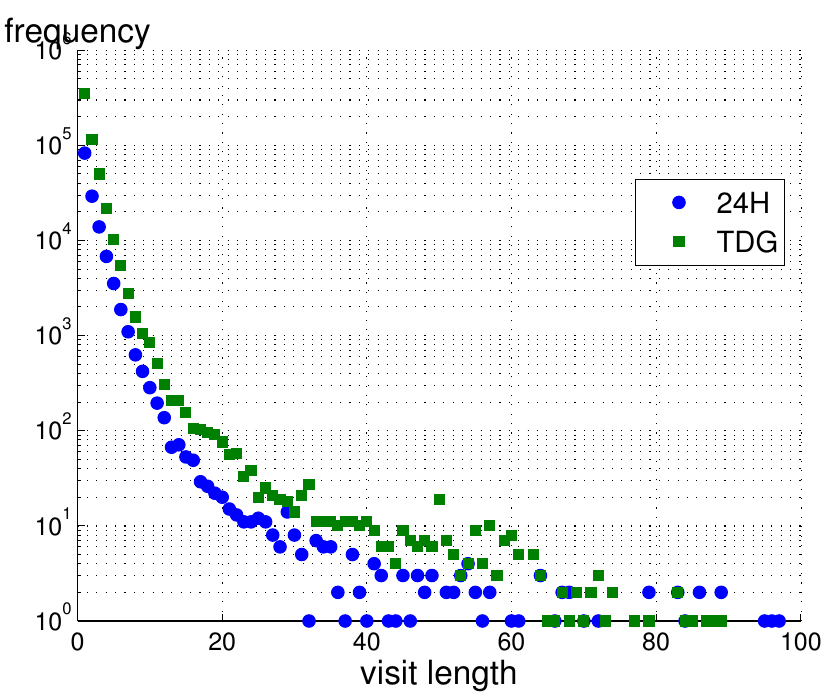}
\caption{Distribution of the length of visits.}
\label{fig:visitDistrib}
\end{figure}

The datasets span from Nov. 2008 until May 2009. They contain all the news stories displayed, and all the visits by anonymous users within the time period. Note that a new visit is created every time a user browses the website, even if she browsed the website before. The raw data has a lot of noise due to, for instance, crawling bots from search engines or browsing on mobile devices with unreliable internet connections. Table \ref{table:dataset} shows the dataset statistics after filtering out this noise, and Figure~\ref{fig:visitDistrib} illustrates the distribution of visit length for each dataset.

\begin{table}[t!]
\centering
\begin{tabular}{lccc}
& \textbf{News stories} & \textbf{Visits} & \textbf{Clicks}\\ \hline\hline
TDG & 10'400 & 600'256 & 1'069'131\\ \hline
24H & 8'613 & 249'099 & 509'978\\
\end{tabular}
\caption{Datasets after filtering.}\label{table:dataset}
\end{table}

\subsection{Evaluation Metrics}\label{sec:evalmetric}

%There has been a lot of discussion on the best way of evaluating recommender systems \cite{Herlocker2004, Shani2011}. The best would be to implement them on an actual site and measure the click rate on recommended items. Unfortunately, this is usually far too costly to do, and evaluation has to be carried out based on behaviour that was observed without the recommender being present. In our case, we have visit histories from the newspaper websites, and we can evaluate how well our recommendations match the news items that readers selected themselves. It is clear that this is a somewhat inaccurate measure: \emph{a)} the user may not have liked all the items she visited; \emph{b)} the user may have preferred one of the recommended items to the one she clicked, so the fact that a recommended item was not visited does not mean the recommendation is bad. However, we believe that prediction of the visit history is still a useful way to \emph{compare} the performance of different techniques, so we use it here.

We evaluate how good the systems are in predicting the future news a user is going to read. Specifically, we consider sequences of news items $\seq = \langle n_1, n_2, ..., n_l \rangle$, $n_i \in \news, \seq \in \CS$ read by anonymous users. The sequences and the news items in each sequence are sorted by increasing order of visit time. When an anonymous user starts to read a news item $n_1$, the system generates 5 recommendations. As soon as the user reads another news item $n_2$, the system updates its model with the past observations $n_1$ and $n_2$, and generates a new set of recommendations. Hence the training set and the testing set are split based on the current time: at time $t$, the training set contains all news items accessed before $t$, and the testing set has items accessed after $t$.

We consider three metrics. The first is the \textbf{Success@5} (s@5). For a given sequence $\seq = \langle n_1, n_2, ..., n_t, ..., n_l \rangle$, a current news item $n_t$ in this sequence, and a set of recommended news items $\mathcal{R}$, s@5 is equal to $1$ if $n_{t + 1} \in \mathcal{R}$, $0$ otherwise.

The second metric is \textbf{personalized s@5}, where we remove the popular items $\mathcal{R_T}$ from the set $\mathcal{R}$, to get a reduced set $\mathcal{R}_P = \mathcal{R} \setminus \mathcal{R}_T$. This metric is important because it filters out the bias due the fact that data is collected from websites which recommend the most popular items by default.

The final metric is \textbf{novelty}, defined by the ratio of unseen and recommended items over the recommended items: $novelty = |\mathcal{R} \cap \mathcal{F}| / |\mathcal{R}|$. This metric is essential because users want to read about something they do not already know.

%We recommend 5 stories, and use the \textbf{Success@5} (s@5) metric to evaluate how good the recommendations are\footnote{For all the following figures, we witnessed the same behaviour with the mean average precision metric. We omit the figures due to space constraint.}. For a given sequence $\seq = \langle n_1, n_2, ..., n_t, ..., n_l \rangle$, a current news item $n_t$ in this sequence, and a set of recommended news items $\mathcal{R}$, s@5 is equal to $1$ if $n_{t + 1} \in \mathcal{R}$, $0$ otherwise.

%Recommender systems should bring added value, and not just recommend the most popular stories that the reader would have already seen on the front page. Hence, we also consider the \textbf{personalized s@5} where we remove the popular items $\mathcal{R_T}$ from the set $\mathcal{R}$, to get a reduced set $\mathcal{R}_P = \mathcal{R} \setminus \mathcal{R}_T$.

%For the news domain, novelty is essential because users want to read about something they do not already know. Therefore in addition to accuracy, we consider \textbf{novelty}. We define the novelty as the ratio of unseen and recommended items over the recommended items: $novelty = |\mathcal{R} \cap \mathcal{F}| / |\mathcal{R}|$.

\subsection{Sensitivity Evaluation}\label{sec:results}

For all systems, we use a prior $\alpha_0 = 1 / |\news|$ for the Dirichlet models, and the initial weights for the experts as $w_k = 2^{-k}$, where $k$ is the depth of the node. We evaluated experimentally the optimal number of topics in the range of 30 to 500, and found that 50 topics bring the best accuracy. We varied the size of candidate pool: number of popular items $|\mathcal{P}|$ from 10 to 500, and fresh items $|\mathcal{F}|$ from 10 to 100. When the candidate set is small, the experts consider only the most recent read stories. We report averages over all recommendations with confidence intervals at 95\%. We omit figures for the TDG dataset because we witnessed the same behaviours.

\begin{figure}[t!]
\centering
\includegraphics[scale=0.5]{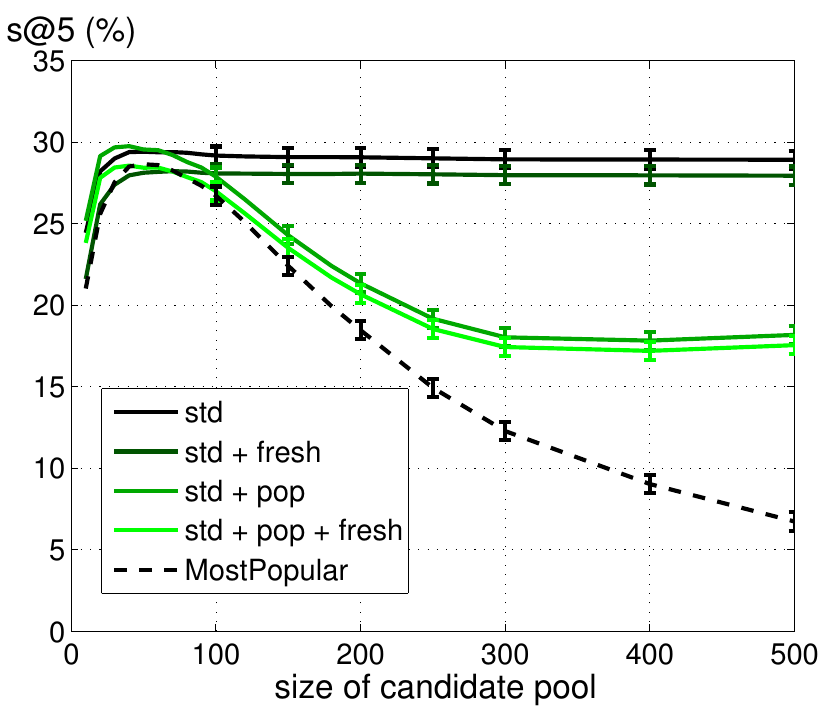}
\caption{VMM recommender system: different mixtures of experts (Bayesian update, $|\mathcal{F}| = 10$).}
\label{fig:VMM_mixture}
\end{figure}

\begin{figure*}[t!]
\centering
\subfigure[Non-personalized news items]{\includegraphics[scale=0.45]{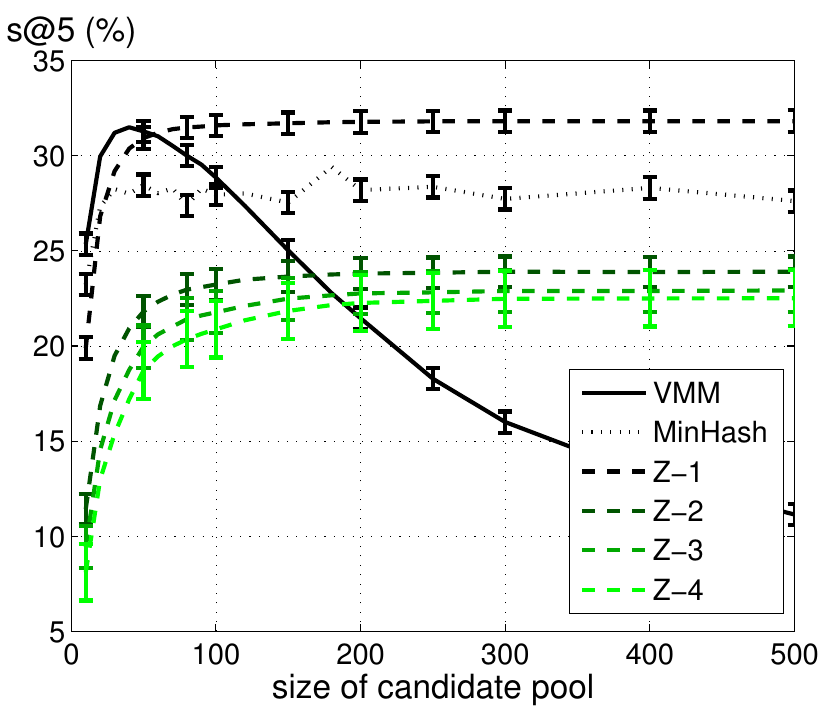}\label{fig:noperso}}
\subfigure[Personalized news items]{\includegraphics[scale=0.45]{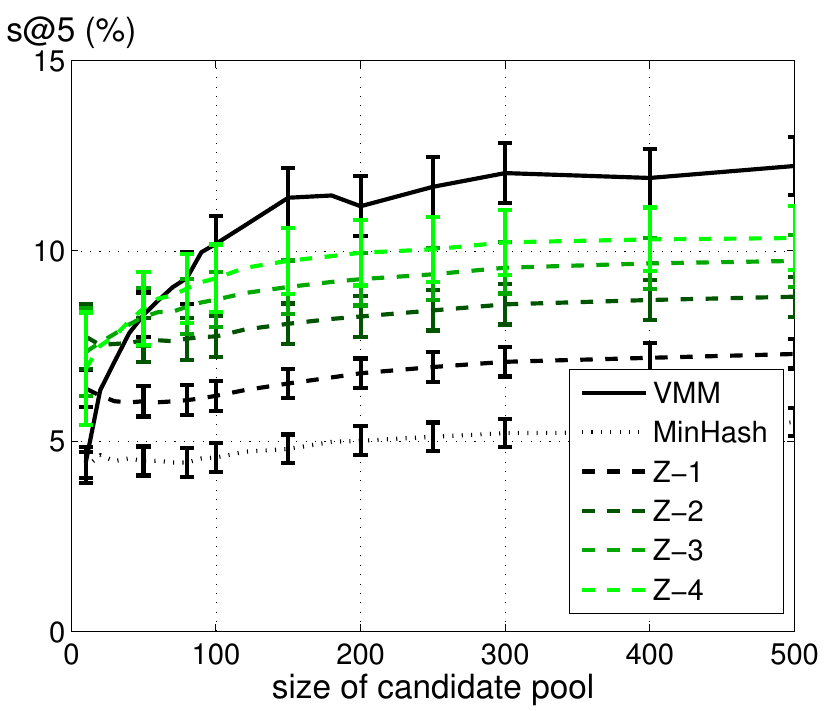}\label{fig:perso}}
\caption{Accuracy for personalized and non-personalized news items.}
\label{fig:allperso}
\end{figure*}

\begin{figure*}[t!]
\centering
\subfigure[Tuning: 24H dataset]{\includegraphics[scale=0.45]{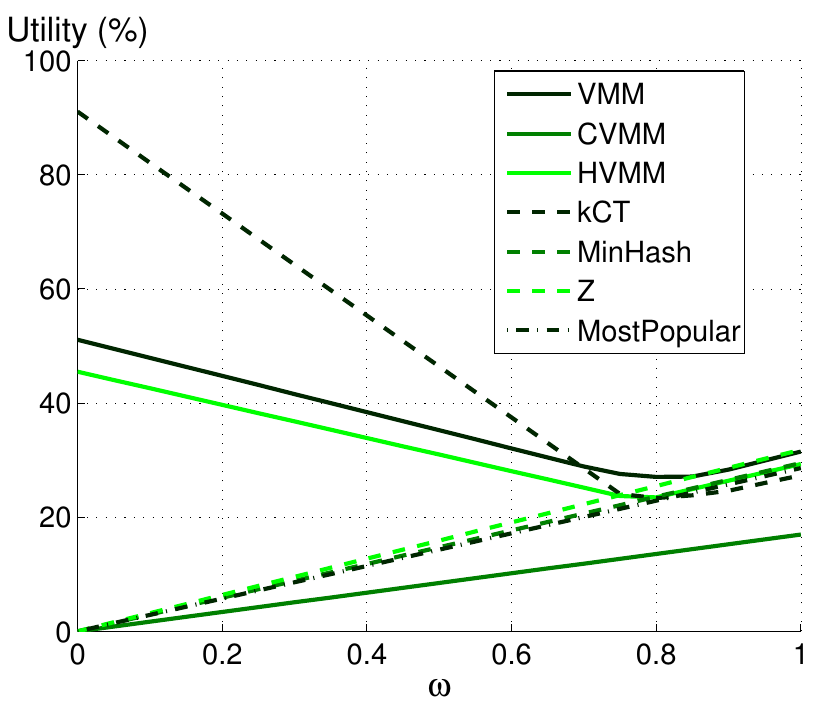}\label{fig:EPC_24h}}
\subfigure[Testing: TDG dataset with optimal parameters from 24H dataset.]{\includegraphics[scale=0.45]{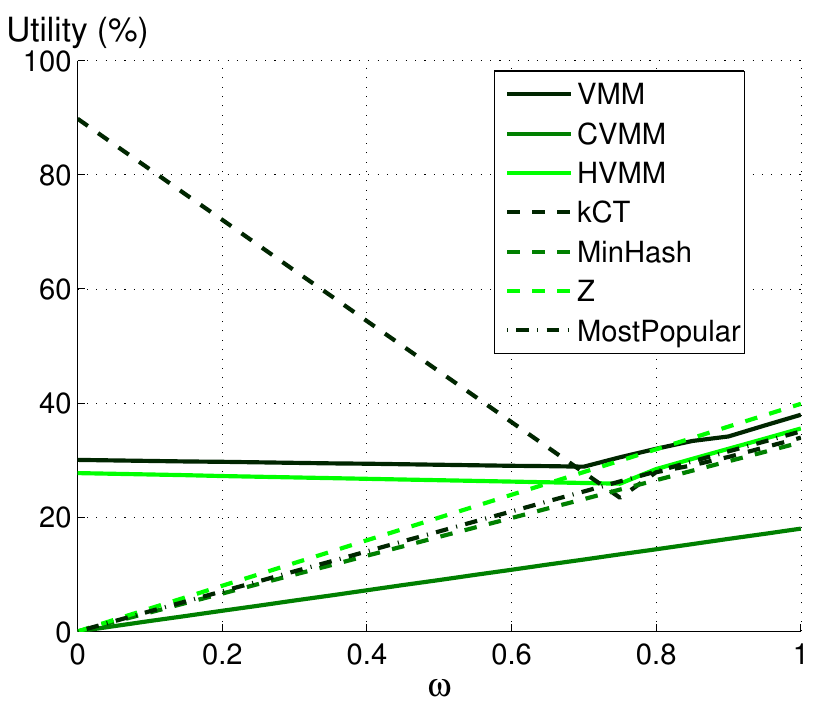}\label{fig:EPC_tdg}}
\caption{Expected performance curves: accuracy and novelty trade-off.}
\label{fig:EPC}
\end{figure*}

Although na\"{\i}ve, the approach of recommending the most popular stories is actually used very often on newspaper websites. This strategy does not pay off when the size of candidate pool increases. ''Good" recommendations are drowned in popular items. This can also be seen by the fact that mixture of expert models integrating the popularity model are very sensitive to the number of popular items while others are more robust (e.g. Fig.~\ref{fig:VMM_mixture} for VMM recsys). 

We noticed that, when using the Dirichlet priors to update the mixture probabilities, the prediction was mostly made by the popularity model, resulting in the same behaviour as the most-popular recommender system as the size of candidate pool increases. However, as the Bayesian update (Eq.~\ref{eq:bayesian}) adapts the probabilities based on the performance of each expert model, it is more robust when we increase the candidate set. We also observed that as the number of fresh items increased, CT models were getting slightly better.

When we look at the general accuracy of CT recommender systems (Fig.~\ref{fig:noperso}), their performance is close to the existing techniques. However when we consider only personalized items (Fig.~\ref{fig:perso}), CT recommender systems outperform current techniques, showing that the order of the model is important. Indeed, we observed that the weights of the experts are well distributed over the space even for long sequences. If the sequence is not important, the weights of the experts would have been 0 except for the root expert, resulting in a performance similar to Z-$1$.

We have seen that recommending popular news is easy and relatively accurate. However, a recommender system with a high accuracy on an offline evaluation does not imply that in practice it will give useful recommendations to the users. In the context of news, novelty plays a crucial role, and the users expect both personalized and novel recommendations. In the next section, we study this trade-off.

\subsection{Comparison}

In practice, we may be interested in some particular mix of accuracy and novelty. We formalize this by defining a utility function: $U(\omega | D, A) = \omega * s@5 + (1 - \omega) * novelty$, where $\omega$ specifies the trade-off between accuracy and novelty, $D$ is the dataset (24H or TDG) and $A$ is an assignment of parameters. For CT systems, the parameters are the number of popular $|\mathcal{P}|$ and fresh $|\mathcal{F}|$ items, whether the probabilities are computed via a Bayesian update or not, the mixture of experts (standard, popularity and/or freshness). It might be the case that different parameters are optimal for different utilities. The same holds for the parameters of the other methods we compare against.

To perform the comparison, we simulate the process of a designer who is going to tune each system on a small dataset (24H), before deploying the recommender online (on the TDG dataset). For any value of $\omega$, we find the best parameters for the 24H dataset, and then measure the performance on the other dataset. This gives the Expected Performance Curve (EPC) \cite{Bengio2005}, which provides an unbiased evaluation of the performance obtained by different methods.

Figure~\ref{fig:EPC} illustrates the EPCs for 24H and TDG datasets. Fig.~\ref{fig:EPC_24h} shows the optimal utility $U(\omega | D, A^*(\omega, D))$ with $A^*(\omega, D) = \arg\max_A U(\omega | D, A)$ for $D = \text{24H}$ dataset. Figure~\ref{fig:EPC_tdg} shows the corresponding utility $U(\omega | D', A^*(\omega, D))$ achieved on the test dataset $D' = \text{TDG}$ using the parameters found for the tuning dataset. First, we observe that all methods are robust in that they have similar performance in the testing dataset. Second, we observe that the purely content-based method (CVMM) performs poorly both with respect to novelty and accuracy. The hybrid approach (HVMM) is significantly better. Third, the approaches that disregard the content (VMM and Z) perform similarly in terms of accuracy, but only the VMM has a reasonable novelty. Finally, the $k$-d tree approach ($k$CT) has a much higher novelty than anything else. Thus, if one were to select a method based on performance on the tuning set, one should choose $k$CT for smaller values of $\omega$ and VMM for larger values.

\section{Conclusion}\label{sec:conclusion}

News recommendation is challenging due to the rapid evolution of topics and preferences. We introduced a class of recommender systems based on context trees that accommodate a dynamically changing model. We considered context trees in the space of sequences of news, sequences of topics, and in the space of topic distributions. We defined expert models which consider the popularity and freshness of news items, and examined ways to combine them into a single model. We proposed an incremental algorithm that adapts the models continuously, and is thus better suited to such a dynamic domain as the context tree evolves over time and adapts itself to current trends and reader preferences. Our approach requires only one tree (the context tree), and thus scales very well. Our work is not restricted to the history of logged-in users, but considers a one-time session for recommendation, where users do not log in. Surprisingly, we do not know of any existing research that considers context-tree models for recommender systems.

Each proposed variant has its strengths and weaknesses. To evaluate them, we used the expected performance curve methodology, whereby each method is tuned in a training set according to a parametrized utility metric. In doing so, we showed that if we are interested in accuracy in a static dataset, a context tree that implements a variable-order Markov model is ideal, while novelty is best served with a $k$-d tree on the space of topics. In addition, we showed that a large order is mainly important when we are not interested in recommending highly popular items. An open question is whether the results we obtained on a static trace will be qualitatively similar in an actual recommender system. In future work, we aim to perform an online test of the system on a real news website.

%We showed that CT recommender systems are flexible enough to capture the properties of news items, with an accuracy that is comparable to the best existing techniques but bringing novel and personalized recommendations.

\bibliographystyle{abbrv}
\bibliography{arxivv2}

\end{document}